\documentclass[preprint,aps]{revtex4-1}
\usepackage{soul} % strike through package (added by Pabitra)

\usepackage{color}
\usepackage{mathtools}
\usepackage{array}
\usepackage{tabularx}
\usepackage{diagbox}
\usepackage{tikz}
\usetikzlibrary{decorations.pathmorphing}
\usepackage{bm}
\usepackage{amsmath,bm,amssymb,amsfonts,dcolumn,color,graphicx,latexsym,epsfig}
\usepackage{rsfso}
\usepackage{hyperref}
\hypersetup{
colorlinks=true,
linkcolor=magenta,
filecolor=magenta,      
urlcolor=blue,
citecolor=blue,
}

\usepackage{booktabs} 

\usepackage{subfig}
\usepackage{graphicx}
\usepackage{float}

\usepackage[absolute]{textpos}
\usepackage{tikz}
\usepackage{amsmath}

\usepackage{multirow}
\usepackage{natbib}

\usepackage{float}

\usepackage{booktabs}
\usepackage{pifont}
\usepackage{mathrsfs}
\usepackage{caption}
\usepackage{caption, threeparttable}
\begin{document}

%\title{Five-Dimensional Wormhole: Shadow and Scalar Quasinormal Mode}
%\title{On the Shadow and Stability of Extra-dimensional Wormholes}
%\title{Influence of Plasma on the Shadow of a Rotating Damour-Solodukhin Wormhole}

\title{Decoding Horizonless Spacetime: Plasma-Induced Features in a Rotating Wormhole Shadow}

\author{Pabitra Gayen and Ratna Koley}
\email{pabitra.rs@presiuniv.ac.in, ratna.physics@presiuniv.ac.in}
\affiliation{Department  of  Physics,  Presidency  University,  86/1 College Street,  Kolkata  700073,  India.}

\begin{abstract} 
We investigate the shadow properties in a recently proposed geometry of a rotating wormhole under realistic astrophysical conditions,  
particularly in the presence of a cold and non-magnetized plasma environment surrounding the 
wormhole throat. Using the Hamilton-Jacobi formalism, we derive the orbit equation under specific plasma density profiles, 
where we consider plasma as dispersive medium and disregard it's influence on the background geometry. % 
The electron density distribution is chosen to preserve a generalized Carter constant. 
We explore the shadow cast by this class of rotating wormhole in the presence of both homogeneous and non-homogeneous plasma as seen by an asymptotic observer. The photon regions are visualized, and the influence of geometric parameters, plasma parameters, and the observer’s inclination angle with the rotation axis on the resulting shadow morphology is analyzed. 
We tried to implement constraints on  the plasma and the geometrical parameters of the wormhole such as the spin parameter and the deviation (from Kerr) parameter in the back drop of recent observational bounds coming from the deviation from circularity of the shadow boundary ($\Delta C$) and  deviation of the average shadow radius from Schwarzschild ($\delta$). The bound on $\Delta C$ is satisfied by the theoretically allowed range of parameters thus not found very
useful to put any constraint, we could impose stringent constraints on the parameters based on the observed value of $\delta$.
By comparing the optical characteristics of the image of these wormholes 
with those of Kerr black holes under analogous plasma conditions, we identify the features that 
could serve as discriminants for similar types of compact objects. 

\end{abstract}
%==================================================================================================================================================================================================================


\maketitle

\section{Introduction}

% Shadow EHT in general
Recent observational breakthroughs -- most notably the imaging of the shadows of M87$^{*}$ 
and SgrA$^{*}$  by EHT -- intensified the study of gravity in extreme environments, stimulating efforts to probe compact objects via their shadow features and examining plasma/accretion flow effects on shadow morphology \cite{EventHorizonTelescope:2019dse, EventHorizonTelescope:2019pgp, EventHorizonTelescope:2022wkp, EventHorizonTelescope:2022xqj, Abramowicz:2002vt, Takahashi:2004xh, Narayan:2008bv, Broderick:2009ph, Broderick:2015tda}.
% Wormhole in general
In this context, Lorentzian wormholes (WH) provide a prolific ground for testing different theories of gravity. The wormholes are constrained by energy conditions in 4D GR, however, their existence remains plausible with ordinary matter in other gravity theories, and the role of exotic matter in cosmic acceleration further accentuates their theoretical significance within the framework of 4D GR. Moreover, the possibility that the strong-field phenomena near supermassive compact objects may be closely mimicked by horizonless wormhole geometries \cite{Combi:2024ehi} left wormholes as viable alternatives for galactic cores — at least until an event horizon is unambiguously detected \cite{Abramowicz:2002vt}. 
Considerable progress has been made in the study of static wormholes, particularly with respect to their traversability \cite{Morris:1988cz}, stability \cite{Gonzalez:2008wd, Bronnikov:2012ch, Armendariz-Picon:2002gjc, Torii:2013xba, Novikov:2009vn, Tsukamoto:2018lsg}, matter violating energy conditions \cite{Morris:1988cz} and the non violating ones \cite{Visser:2003yf, Kar:2015lma, Bronnikov:2016xyp, Bronnikov:2018uje, Kuhfittig:2018voi, Blazquez-Salcedo:2020czn, Konoplya:2021hsm}. 
% Rotating WH
According to recent findings a greater stability of the WH solutions might be achieved either through a generalized theory of gravity \cite{Dzhunushaliev:2013jja} or, alternatively, by considering rotating wormholes \cite{Azad:2023iju, Cisterna:2023uqf}. 
The first such attempt to construct a rotating WH was made by Teo \cite{Teo:1998dp}, 
who focused on its general form and properties and this has been further explored in \cite{Shaikh:2018kfv, Gyulchev:2018fmd, Abdulxamidov:2022ofi, Kumar:2023wfp}. The rotating WH as a solution of GR was first reported in \cite{Kleihaus:2014dla}, through an extension of Ellis WH with a phantom scalar and angular momentum, 
which has since motivated extensive research.  
In this paper, we aim to advance the study of rotating wormholes within a more astrophysically realistic framework in the background of a newly proposed rotating WH \cite{Kar:2024ctd} (which can be classified as a Kerr-like WH) by investigating the photon rings, and special features in shadow morphology. \\

% BH Shadow Review
Theoretical inception of the shadow cast by a gravitationally intense star dates back to Synge \cite{synge1966escape} and Luminet \cite{luminet1979image} and that 
%in 1950.  
for a rotating black hole to Bardeen \cite{dewitt1973black, chandrasekhar1998mathematical}.  
Kerr black hole, being astrophysically the most relevant has garnered significant attention in the theoretical and observational community.  
%
 %WH Shadow Review
 Conversely, wormhole shadows extend the concept of black hole silhouettes to horizon less spacetimes, offering potential observational signatures of exotic topology.  Early work on the static wormhole shadows demonstrated photon regions which can generate bright rings and central darkness akin to Schwarzschild black hole even in the absence of an event horizon. However, clear discriminants are also investigated -- a markedly different brightness contrast between the interior and exterior ring of Ellis wormhole shadow \cite{Ohgami:2015nra},
 a cuspy structure of the shadow edge \cite{Gyulchev:2018fmd},
 the crucial role of a rotating wormhole throat in a shadow \cite{Shaikh:2018kfv},
 the frame-dragging twists in photon trajectories introducing asymmetry and potential cusps in the shadow outline,
 splitting of spherical orbits into distinct families whose superposition shapes the rim of the shadow \cite{Gyulchev:2018fmd, Shaikh:2018kfv}, 
 distinctive shadow distortions and shifts in ring-radius depending on the charge parameter for a rotating wormhole threaded by a global monopole charge \cite{Zheleva:2025iib}. \\

%\textbf{Why plasma?} 
%
 Accounting for plasma is also essential in the modeling of wormhole shadows, as realistic astrophysical environments are accretion rich. 
    Plasma alters photon trajectories through refraction and frequency dependence, thereby reshaping the observable shadow structure via chromatic distortions. Since present instruments operate in radio bands where plasma effects are pronounced, their inclusion is essential for capturing frequency-dependent observational signatures specific to wormholes \cite{perlick2017light, Abdujabbarov:2016efm,  Bisnovatyi-Kogan:2015dxa, Rogers:2015dla}. 
Muhleman et al. \cite{Muhleman:1970zz, muhleman1966radio} modeled light propagation through a non-magnetized, pressure less plasma within linearized gravity, while gravitational and plasma deflections are analyzed separately in \cite{1989KiIND....Q....B}; subsequent extensions considering various plasma distributions are found in \cite{perlickray}. 
%% SHADOW WITH PLASMA REVIEW
In the background of wormhole geometries, the plasma around Teo-type WHs reduces the inner radii of circular photon orbits and distorts shadows based on spin, plasma density and observer inclination \cite{Abdujabbarov:2016efm, Alloqulov:2024olb}. Analytical shadow boundaries in non-homogeneous plasma, demonstrating shadow shrinkage up to disappearance due to 
higher and higher plasma densities has been shown in \cite{Kumar:2023wfp}. These works underscore the importance of plasma in defining the features of wormhole shadow. 
%\textbf{Methodology \& Summary of work done}
In this work we further explore light propagation, in a newly proposed rotating wormhole geometry \cite{Kar:2024ctd} in the presence of non-magnetized and pressure less plasma, employing realistic plasma models. Analytical investigation is done by applying the Hamilton-Jacobi formalism in which light rays admit the Carter constant alongside symmetry-originated constants of motion.
%% Plasma modelling
Plasma is modeled considering both homogeneous and non-homogeneous density profiles in such a way that it satisfies the separability condition. Our analysis shows that the interplay of plasma and rotation parameters produces asymmetric, frequency-dependent shadow.  \\

The paper is organized in the following manner. In section II, we re-analyze the rotating wormhole geometry highlighting its special features on the null orbits in the presence of plasma. In the next section III, we work on Hamilton-Jacobi equation in the above mentioned rotating wormhole background. Guaranteeing the separability we find the equation of motion for the light rays in the presence of plasma. Further we determine the spacetime regions filled with spherical light rays for various choices of plasma density profiles. In the following section IV, we study the shadow profile where the knowledge of the constants of motion and the associated conserved quantities play very crucial role in deriving the boundary curve of the shadow on the sky of 
an observer situated at a location away from the throat of the wormhole on one side of the spacetime. 
%Units
In all the calculations we have used geometrical unit choosing $c = G = 1$. Throughout the paper, mostly
positive sign convention has been used for the metric.

\section{Rotating WH Background and Hamiltonian Formalism in plasma environment}

Let us begin by revisiting a recently proposed rotating Lorentzian wormhole spacetime \cite{Kar:2024ctd} which was derived by using the method of Azreg-A\"{\i}nou \cite{Azreg-Ainou:2014pra} applied to a known static, spherically symmetric, non-vacuum wormhole configuration with a vanishing Ricci scalar \cite{Dadhich:2001fu}. In contrast to general relativity, where static configurations typically demand energy-condition violations, certain braneworld models 
exhibit the same static geometry satisfying the null energy condition \cite{Kar:2015lma}. It is noteworthy that the rotating spacetime remains regular and exhibits wormhole characteristics. The explicit form of the rotating wormhole metric in Boyer–Lindquist coordinates is:

\begin{eqnarray}\label{rotating wh}
ds^2 =&-&\left(1-\frac{2m(r)r}{\Sigma}\right)dt^2 - \frac{4m(r)ra \sin^2\theta}{\Sigma}dtd\phi+\frac{\Sigma}{\Delta}dr^2+\Sigma d\theta^2 \nonumber \\
&+& \sin^2\theta \left(r^2+a^2+\frac{2m(r)ra^2\sin^2\theta}{\Sigma}\right)d\phi^2
\end{eqnarray}
where
\begin{eqnarray}
m(r)&=&\frac{r}{2}\left(1-\frac{\left(q+\sqrt{1-2M/r}\right)^2}{\left(q+1\right)^2}\right),  \\
\Sigma(r, \theta) &=& r^2+a^2\cos^2\theta, \\
 \Delta(r)&=&\left(1-\frac{2M}{r}\right)\left(\frac{\left(q+\sqrt{1-2M/r}\right)^2r^2+a^2\left(q+1\right)^2}{\left(q+\sqrt{1-2M/r}\right)^2}\right)
\end{eqnarray}

The rotating wormhole metric in Eq. (\ref{rotating wh}) can be rewritten in the following generic form
\begin{equation}
    \mathrm{d}s^2 = -N^2 dt^2 + \frac{dr^2}{1 - b/r} + r^2 \tilde{K}^2 d\theta^2 + r^2 K^2 \sin^2 \theta \, \left(d\phi - \omega dt\right)^2
\end{equation}
with,
\begin{eqnarray}
N^2(r, \theta) &=& \frac{\Sigma(r^2 + a^2 - 2m(r)r)}{\Sigma(r^2 + a^2) + 2m(r)ra^2 \sin^2 \theta} \nonumber \\
b(r, \theta) &=& r - \frac{r\Delta(r)}{r^2 + a^2 \cos^2 \theta} \nonumber \\
\omega(r, \theta) &=& \frac{ 2m(r)ra}{\Sigma(r^2 + a^2) +2m(r)ra^2 \sin^2 \theta}  \\
\tilde{K}^2(r, \theta) &=& 1 + \frac{a^2 \cos^2 \theta}{r^2} \nonumber \\
\mathcal{R}^2(r, \theta) &=& r^2 K^2(r, \theta) = r^2 + a^2 + \frac{2m(r)ra^2 \sin^2 \theta} {\Sigma} \nonumber
\end{eqnarray}
The spacetime geometry is characterized by three parameters: (i) $M$, representing the ADM mass; (ii) $a$, the spin parameter; and (iii) $q$, the dimensionless parameter that quantifies deviation from the Kerr solution.  
In the limit  $q \rightarrow 0$, the solution smoothly reduces to the Kerr black hole. Conversely, in the limit $a \rightarrow 0$ rotation gets eliminated and yields a static configuration \cite{Kar:2015lma} and in the asymptotic limit the spacetime becomes flat. 
It is important to emphasize that this geometry does not correspond to a Teo-type rotating wormhole, but rather resembles a Kerr-like geometry—serving as the rotating extension of the Damour–Solodukhin wormhole \cite{Bueno:2017hyj, Damour:2007ap}. 
A distinctive feature of this spacetime is the absence of an ergosphere since the $g_{tt}$~component of the metric remains negative for all $r$ and $\theta$ values. An analysis of the Zakhary–McIntosh invariants reveals that, although the Ricci scalar is non-zero for finite values of $a$ and $q$, all curvature invariants remain finite—thereby confirming the absence of singularities in the rotating wormhole spacetime. 
%The radius of the 
%throat depends on both $a, q$ and $M$: $$S|_{r = 2M} = \sqrt{4M^2 + \frac{a^2(q^2 + 4q+2)}%{(q^2+2q+1)}}.$$ 

The exotic matter required to uphold this rotating wormhole structure violates all classical energy conditions, assuming the geometry is treated as a solution within the framework of GR. In braneworld scenarios, the rotating wormhole may evade the need for exotic matter, offering a more viable configuration than in standard 4D general relativity. 
In the subsequent section, we undertake a detailed analysis of photon trajectories which incorporates the influence of a non-magnetized, pressure less, dispersive plasma medium, which modifies the null geodesics due to its frequency-dependence. 

There is good reason to assume that massive compact objects are embedded in ionized plasma and surrounded by accretion disks. Such environment is treated as a dispersive, pressure less, and non-magnetized plasma medium which affects light propagation in a frequency-dependent manner, causing them to deviate from geodesics particularly in the radio frequency range. The Hamiltonian for light rays in a plasma can be derived from Maxwell’s equations, where the electromagnetic field arises from two charged fluids - ions and electrons. A simplified derivation in curved background for the non-magnetized, pressure less plasma is done in \cite{perlickray}. 
In contrast to vacuum the propagation of light in a plasma environment is described by the modified Hamiltonian 
\begin{equation}\label{gen hamiltonian}
    H(x,p)=\frac{1}{2}\left\{g^{\mu \nu}(x)p_\mu p_\nu+\omega_P(x)^2\right\}
\end{equation}
where $x$ stands for $(t,r,\theta,\phi)$ and $p$ for $(p_t,p_r,p_\theta,p_\phi)$. 
The plasma density profile is typically described using the Drude-Lorentz model, where the plasma frequency $\omega_P$ at position $x$ is proportional to $\sqrt{N_e(x)}$, the square root of the local electron number density
\begin{equation}\label{plasma frequency}
\omega_P(x)^2=\frac{4\pi e^2}{m_e}N_e(x)
\end{equation}
where $e$ and $m_e$ are the charge and mass of electron respectively.  The observed frequency of light rays, $\omega_{l}(x)$, is gravitationally red shifted and it can be expressed in terms of the conserved momentum \( p_t \) as 
\begin{equation}\label{redshifted frequency}
\omega_{l}(x)=\frac{p_t}{\sqrt{-g_{tt}(x)}}.
\end{equation}
Recalling that the phase velocity of a particle is given by the ratio of energy and linear momentum, the frequency dependent refractive index of the plasma medium as measured by a local timelike observer is given by the relation
\begin{equation}\label{refractive index1}
n\left(x,\omega_{l}(x)\right) = \sqrt{1-\frac{\omega_P(x)^2}{\omega_{l}(x)^2}}, ~~~~~ \omega_{l}^2 \geq {\omega_P}^2.
\end{equation}
The frequency of photon must satisfy the above inequality so as to reach the asymptotic observer without getting absorbed by the medium. 
Heuristically, plasma acts like a refractive lens, magnifying or demagnifying the shadow depending on the specific model. 
For the rotating wormhole spacetime metric in Eq. (\ref{rotating wh}) the Hamiltonian (in Eq. \ref{gen hamiltonian}) reduces to 
\begin{eqnarray}\label{hamiltonian for rot wh}
H= \frac{1}{2\Sigma}\left[\left\{\frac{1}{\sin^2\theta}-\frac{a^2}{Q(r)}\right\}p_\phi^2 + \left\{a^2\sin^2\theta-\frac{\left(r^2+a^2\right)^2}{Q(r)}\right\}p_t^2 \right] \\ \nonumber 
+ \frac{1}{2\Sigma}\left[2a\left\{1-\frac{r^2+a^2}{Q(r)}\right\}p_tp_\phi + \Delta \left(r\right) p_r^2+p_\theta^2+\Sigma \omega_P^2\right]
\end{eqnarray}
where $Q(r)=r^2+a^2-2m(r)r$.  As the metric in eq. (\ref{rotating wh}) is stationary and axially symmetric, there is a time-translation symmetry leading to a timelike Killing vector, $\xi^\mu = \left( {\partial}/{\partial t} \right)^\mu$ and a rotational symmetry about the axis of rotation leading to a spacelike Killing vector, $\psi^\mu = \left({\partial}/{\partial \phi} \right)^\mu$. The associated conserved quantities are $p_t = - \omega_0 \text{(energy)}$ and \(p_\phi = L\) \text{(angular momentum)} respectively. 
These symmetries are crucial as they allow for separation of variables in the Hamilton-Jacobi equation, enabling ray tracing and shadow analysis. Let us now choose the following separation ansatz  for the Jacobi action:  
\begin{equation}\label{separation ansatz}
S\left(t,r,\theta,\phi \right)=p_tt+p_\phi \phi +S_r\left(r\right)+S_\theta \left(\theta\right)
\end{equation}
where $S_r\left(r\right)$ \& $S_\theta \left(\theta\right)$ are functions of only $r$ and only $\theta$ respectively.
The above ansatz along with the conserved quantities $p_t$ and $p_{\phi}$, yields the following form of the Hamilton-Jacobi equation: 
\begin{eqnarray}\label{hamilton jacobi eq3}
\left\{\frac{1}{\sin^2\theta}-\frac{a^2}{Q(r)}\right\}L^2&+&\left\{a^2\sin^2\theta-\frac{\left(r^2+a^2\right)^2}{Q(r)}\right\}\omega_0^2-2a\omega_0 L\left\{1-\frac{r^2+a^2}{Q(r)}\right\} \\ \nonumber 
&+& \Delta \left(r\right) \left(\frac{dS_r}{dr}\right)^2+\left(\frac{dS_\theta}{d\theta}\right)^2+\Sigma \omega_P^2=0
\end{eqnarray}
The applicability of the Hamilton–Jacobi formalism requires the separability of the underlying equation, which can be achieved only for specific classes of plasma distributions. To address this, we invoke the necessary and sufficient condition for separability established in \cite{perlick2017light} which prescribes that the plasma frequency must take the following form  
\begin{equation}\label{gen plasma frequency}
\omega_P^2(r,\theta)=\frac{f_r(r)+f_\theta(\theta)}{r^2 + a^2 \cos^2 \theta}
\end{equation}
where \( f_r(r) \) denotes an arbitrary function of the radial coordinate $r$ and \( f_\theta(\theta) \) 
represents an arbitrary function of the angular coordinate $\theta$ alone.  
Applying this we can rearrange the Hamilton-Jacobi equation to achieve
\begin{eqnarray}
&-&\Delta\left(r\right)\left(\frac{dS_r}{dr}\right)^2+2a\omega_0 L\left\{1-\frac{r^2+a^2}{Q(r)}\right\}-\left\{a^2-\frac{\left(r^2+a^2\right)^2}{Q(r)}\right\}\omega_0^2-\left\{1-\frac{a^2}{Q(r)}\right\}L^2-f_r\left(r\right) \nonumber \\
&=& \left(\frac{dS_\theta}{d\theta}\right)^2+L^2\cot^2\theta-a^2\omega_0^2\cos^2\theta+f_\theta\left(\theta\right)
= C
\end{eqnarray}

It is crucial to note that the profile in Eq.(\ref{gen plasma frequency}) results in another associated constant of motion  - the Carter constant, $C$ which makes the equation for light rays completely integrable. 
Let us now concentrate on the radial equation and the associated effective potential with an aim to study the photon rings and shadow features in the presence of plasma
\begin{equation}\label{radial eq}
\Delta\left(r\right)Q\left(r\right)\left(\frac{dS_r}{dr}\right)^2=-V\left(r\right)
\end{equation}
where the effective potential, $V(r)$ is given by 
\begin{equation}\label{effective potential}
V\left(r\right)=-\omega_0^2\left\{\left(r^2+a^2-a\eta\right)^2-\xi Q(r)-\frac{1}{\omega_0^2}f_r(r)Q(r)\right\}
\end{equation}

 Photon trajectories are physically admissible for the effective potential satisfying the relation $V(r) \leq 0$. Note that, $\eta=L/\omega_0$ and $\xi=K/\omega_0^2$ and $K=C+(L-a\omega_0)^2 >0$ in the above expression. 
\section{Light propagation in specific plasma environments}
Let us consider three distinct representative plasma distribution profiles -- (i) homogeneous, (ii) longitudinal, and (iii) radial for further study. A necessary criterion for choosing these plasma profiles is that they fulfill the separability condition outlined in Eq. (\ref{gen plasma frequency}) and admit a generalized Carter constant. In the following we discuss some plasma density profiles that satisfy the separability condition.  \\

(i) \textit{Homogeneous plasma:} The plasma is homogeneously distributed and is characterized by the following density profile 
\begin{equation}\label{constant plasma}
    \frac{\omega^2_P}{\omega^2_0}=\rho
\end{equation}

where the parameter $\rho$ represents a dimensionless quantity linked to the homogeneous distribution \cite{perlick2017light}. The physical constant of motion $\omega_0$ plays an important role in determining the path of a light ray. In the presence of plasma if $\omega_0$ becomes small compared to $\omega_P$ it may become 
impossible for light rays to propagate through plasma, however, in vacuum there is no such restriction. For any real value of $\omega_0$ the validity condition is satisfied. Note that the homogeneous plasma distribution satisfies the separability condition with 

\begin{equation}\label{constant plasma2}
f_r(r)=\rho \omega_0^2r^2,\quad f_\theta(\theta)=\rho \omega_0^2a^2\cos^2\theta
\end{equation}

%In this case the curvature parameters are affinely related to the proper time of the metric.  \\

(ii) \textit{Longitudinal plasma}: Unlike the homogeneous case, this distribution assumes a monotonic decrease in plasma density with increasing radial coordinate $r$. The analytical form of such distribution was first conceived by Shapiro \cite{shapiro1974accretion}. We proceed by considering a plasma distribution that varies with the polar angle $\theta$ by $f_{\theta} (\theta)$ and $f_r(r) = 0$ such that the separability condition is satisfied. The resulting plasma distribution is expressed as follows

\begin{equation}\label{plasma dist2}
\frac{\omega^2_P}{\omega^2_0}=\frac{\rho_\theta M^2(1+2\sin^2\theta)}{r^2+a^2\cos^2\theta}
\end{equation}

 where $\rho_{\theta}$ represents the longitudinal plasma parameter and it satisfies the condition $\rho_\theta \geq 0$. From the Eqs. (\ref{gen plasma frequency}) and (\ref{plasma dist2}), the plasma profile is described by
\begin{equation} \label{plasma dist2 function}
f_r(r)=0, \quad  f_\theta(\theta)=\rho_\theta \omega_0^2M^2(1+2\sin^2\theta)   
\end{equation}

The separability condition demands $f'(r) = 0$, therefore any constant $f_r$ may be incorporated with 
the profile, we choose the constant to be zero without loss of generality. \\

(iii) \textit{Radial plasma profile:} In the next example we consider a 
non-homogeneous profile which impact the light propagation by a radial variation in the density. For the rotating wormhole metric given by Eq. (\ref{rotating wh}), the plasma distribution is expressed as follows

\begin{equation}\label{plasma dist3}
\frac{\omega^2_P}{\omega^2_0}=\frac{\rho_r \sqrt{M^3r}}{r^2+a^2\cos^2\theta}
\end{equation}

The simplest choice of plasma density proportional to $r^{-3/2}$ \cite{shapiro1974accretion} will not work in the rotating scenario because the separability condition is not satisfied for $a \neq 0$. Thus a modification is introduced by including the dependence on the polar angle $\theta$. 
The radial plasma density parameter $\rho_r$ is subject to the constraint $\rho_r \geq 0$ and the eqs.  (\ref{gen plasma frequency}) and (\ref{plasma dist3}) yield 
\begin{equation} \label{plasma dist3 functions}
f_r(r)=\rho_r \omega_0^2\sqrt{M^3r}, \quad f_\theta(\theta)=0 
\end{equation}

\smallskip

\begin{figure}[h]
         \includegraphics[width=5.3 cm, height = 3.2 cm]{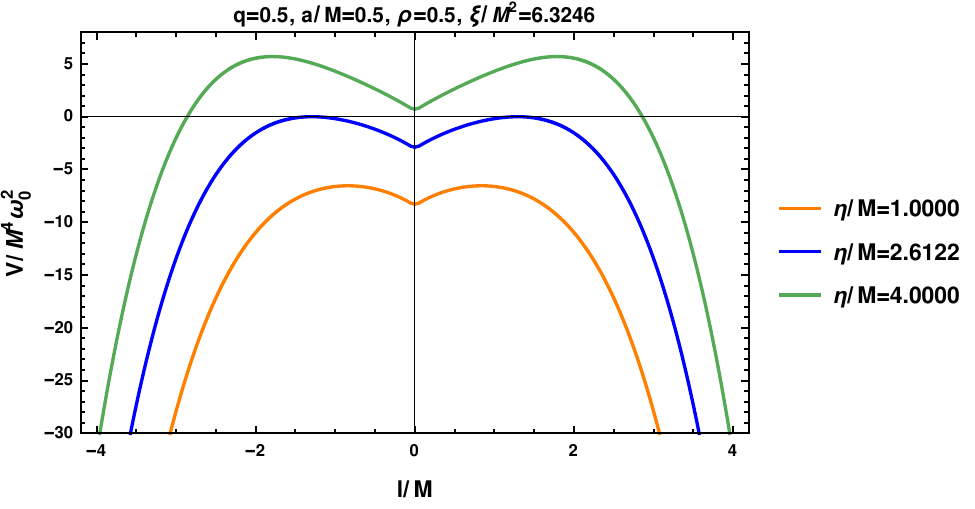}~~~
         \includegraphics[width=5.5 cm, height=3.2 cm]{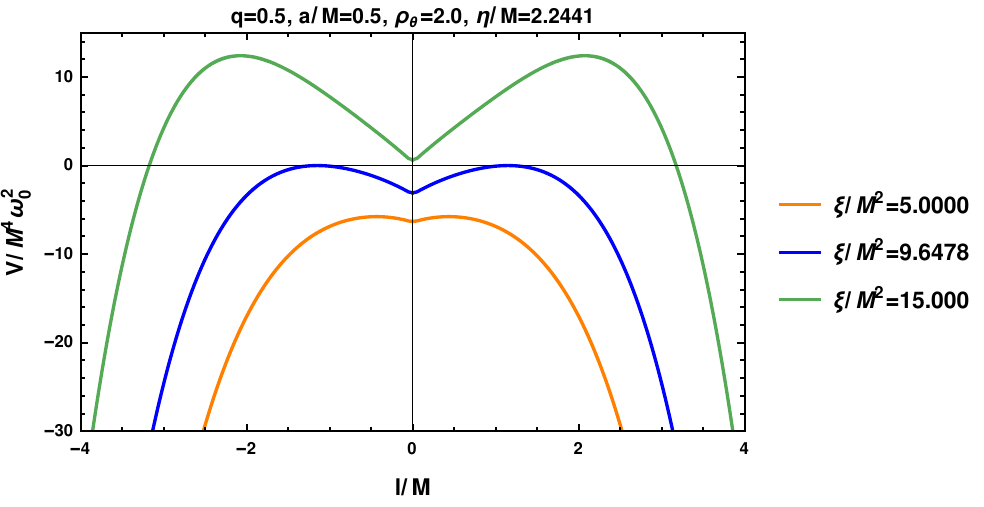}~~~
         \includegraphics[width= 5 cm, height = 3.2 cm]{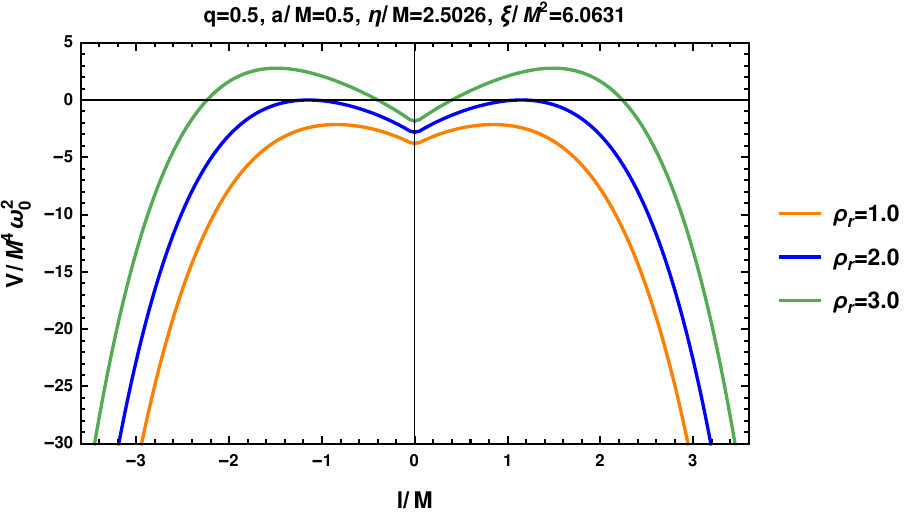}
        \caption{\textit{Representing the behavior of effective potential with $l(r)$ in various situations.}}
        \label{fig:effpot}
\end{figure}

For the construction of the shadow we consider all past oriented light rays emanating from a designated observer position and assign positive frequency to these rays. The photon region is full with spherical light rays that stay on a $r = constant$ sphere in the Boyer-Lindquist coordinate. Before indulging into finding the appropriate spherical rays let us briefly analyze the effective potential and the impact of the associated parameters in determining the photon paths.
In order to envisage the effective potential we engage the proper radial distance as
\begin{equation}\label{proper radial dist}
l(r) = \pm \int_{2M}^{r} \frac{dr}{\sqrt{\Delta/\Sigma}},
\end{equation}

where the $\pm$ signs correspond to the two opposite asymptotically flat regions (Region I and II respectively) connected by the wormhole throat located at $l(2M) = 0$. 
To study the null trajectories we solve the orbit equation numerically and to have an intuitive idea we plot the effective potential in figure \ref{fig:effpot}. The spin parameter, $a$ and deviation parameter $q$ are set at intermediate values to capture optimized impact. 
In the left panel we consider the homogeneous plasma distribution to showcase the possible natures
of the effective potential for different configurations of symmetry specified constants of motion, $\eta$ and $\xi$. Keeping $\xi$ fixed and varying the impact parameter $\eta$, the potential profiles reveal three characteristic photon trajectories depending on the relative strength of wormhole geometrical parameters, plasma density parameter and $\eta$ \& $\xi$ -- (i) a deflected trajectory (green curve), (ii) unstable circular orbit (blue curve), and (iii) a throat-crossing trajectory linking the two regions across the wormhole (orange curve). 
Upon analyzing the alternative plasma profiles and other combinations of impact parameters, we observe analogous behavioral patterns in the potential. In the middle panel we have chosen longitudinal plasma profile and varied $\xi$ while keeping $\eta$ fixed and the right panel presents variations of the effective potential with plasma parameters for radial distribution, keeping both impact parameters and the geometrical parameters fixed. The blue curves give the most interesting feature corresponding to the shadow as these provides the light rings. 

We now explore the photon trajectories and their various shapes based on the choices of geometrical parameters and plasma profiles. 
In spherically symmetric backgrounds the photon ring is the projection of photon sphere on the equatorial plane ( \textit{i.e.}, $\theta=\tfrac{\pi}{2}$ with $\dot{\theta}=0$) and we obtain the photon orbit equation as given below, by using eqs. (\ref{hamiltonian for rot wh}) and (\ref{radial eq}) 

\begin{equation}\label{orbit eq}
    \left(\frac{dr}{d\phi}\right)^2= \frac{Q \Delta  \left\{\left(r^2+a^2-a\eta\right)^2-\xi Q-\frac{1}{\omega_0^2}f_rQ\right\}}{\left\{\left(r^2+a^2-Q\right)a+\left(Q-a^2\right)\eta \right\}^2}
\end{equation}

\begin{figure}[htb]
         \includegraphics[width=0.33\textwidth]{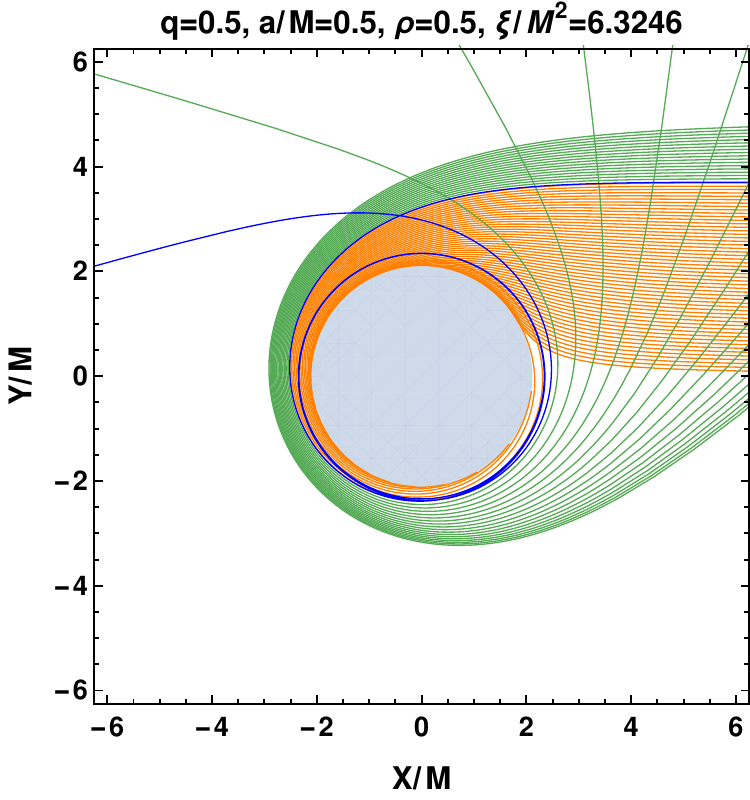}~~~
         \includegraphics[width=0.33\textwidth]{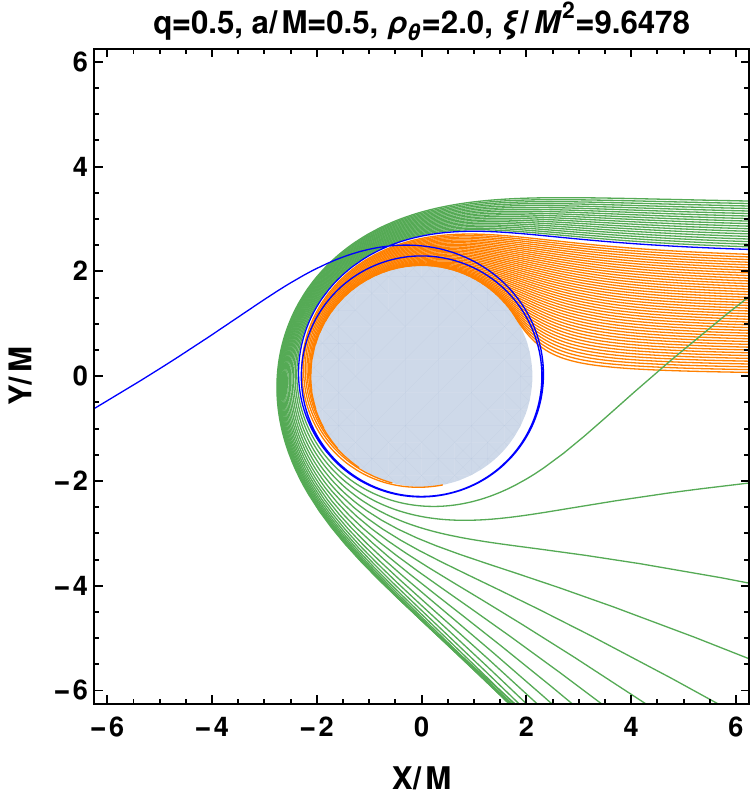}~~
          \includegraphics[width=0.33\textwidth]{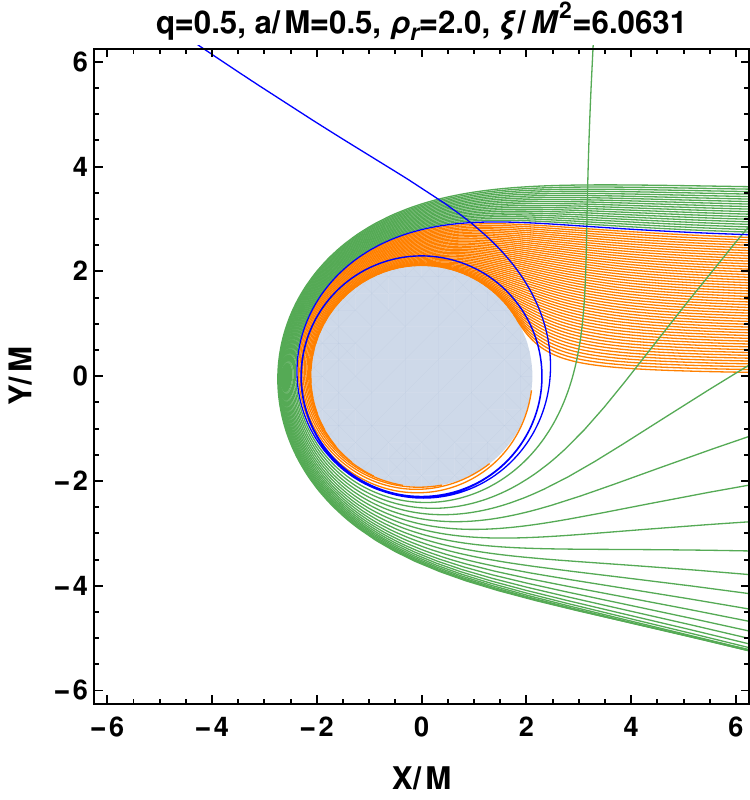}
        \caption{\textit{Photon orbits shown with $X = \mathcal{R}\cos(\phi)$ and $Y = \mathcal{R}\sin(\phi)$. 
    Left panel corresponds to homogeneous plasma, fixed $\xi$ and varying $\eta$. 
        Photon orbits in longitudinal plasma environment is shown in middle panel.
        Right panel depicts photon orbits in radial plasma environment. 
    Orange curves correspond to throat-crossing trajectories, the blue indicates unstable circular orbit, and green curves represent scattered photon paths.}}
        \label{orbits}
\end{figure}

The solution of equation (\ref{orbit eq}) is obtained numerically to study the photon orbits in the rotating wormhole spacetime and we illustrate the trajectories in a systematic fashion in figures (\ref{orbits}) in the presence of homogeneous, longitudinal, and radial plasma distributions  respectively. In each figure, the bluish-gray central area corresponds to the wormhole throat region. \\

For the given background geometry and specified plasma characteristics there exists a combination of critical
parameters $(\xi_c,\eta_c)$ corresponding to an unstable circular orbit (blue curve). In the left panel of Fig. (\ref{orbits}) we have chosen homogeneous plasma to show the effect of varying $\eta$ at fixed $\xi$ and found that the photons with (i) $\eta < \eta_c$ can travel across the throat from one asymptotic region (I) to the other asymptotic region (II) (orange curves), while (ii) those with $\eta > \eta_c$ represent an interesting feature by depicting two turning points, each on one side of the throat. A light ray approaching the throat from one side will thus get scattered to infinity within the same region (green curves) and never reach the throat. The blue curves correspond to the critical value
of the parameter, which have a potential turning point at the photon ring. As these orbits are unstable against perturbation, any light coming from the region I will either scatter back to same side or travel across. Same applies to the light rays coming from the other asymptotic region. Notably, there also exists a critical plasma density parameter, $\rho_c$ associated with the unstable orbit for a given set of $\eta$ and $\xi$. For $\rho > \rho_c$, photons follow scattered trajectories, whereas for $\rho < \rho_c$ they may cross the throat. In the middle and right panels of Fig. (\ref{orbits}) we have shown the orbits in longitudinal and radial plasma environment respectively. We also explored the embedding diagram to see that in this class of wormhole spacetime, photons approaching the throat from region I fall into three classes in the presence of a radial plasma environment for different values of the impact parameters.

\section{Shadow profiles in specific plasma environments}\label{shadow study}
With the aim of finding observational relevance, in this section we determine the shadow cast by the rotating wormhole in three different plasma environments. We choose to study a configuration where the observer is located at an asymptotically flat zone in region I and the wormhole is illuminated solely by light sources originating from region I,  whereas the other region is devoid of any illuminations near the throat under our assumption \cite{Nedkova:2013msa}. Depending on the characterizing parameters the photons  may either (i) enter the wormhole throat and traverse through it, or (ii) be scattered back to infinity. A distant observer detects only the scattered photons, which appear as bright points in the sky, while photons captured by the wormhole correspond to dark spots. This dark silhouette against a bright background formed due to gravitational bending of light constitutes the shadow. The photon orbits are characterized by specific critical values of the impact parameters $\eta$ and $\xi$, such that small perturbations in these values switch the trajectory between capture and scattering. 

The critical values of the impact parameters correspond to unstable circular photon orbits at radius $r_c$ featuring the maximum of the effective potential and is obtained by the following criterion imposed on the potential 

\begin{equation}\label{unstable orbit condition}
V\left(r_c\right)=0, \quad \quad V^\prime \left(r_c\right)=0 \quad \quad \text{and} \quad \quad V^{\prime \prime}\left(r_c\right)<0.
\end{equation}

Hence  the impact parameters $\xi$ and $\eta$ are calculated by using the above conditions (\ref{unstable orbit condition})  as follows

\begin{equation}\label{eta}
\eta=\frac{1}{a}\left[r_c^2+a^2-\frac{Q(r_c)\left\{\sqrt{4 r_c^2 \omega _0^2-f'_r(r_c) Q'(r_c)}+2 r_c \omega _0\right\}}{\omega _0 Q'(r_c)}\right]
\end{equation}

\begin{equation}\label{xi}
\xi=\frac{1}{\omega _0^2 Q'(r_c)^2}\left[Q(r_c)\left\{4 r_c \omega _0 \sqrt{4 r_c^2 \omega _0^2-f'_r(r_c) Q'(r_c)}+ 8 r_c^2 \omega _0^2-f'_r(r_c) Q'(r_c)\right\}-f_r(r_c) Q'(r_c)^2\right]
\end{equation}

These two parameters in eqs.(\ref{eta}) and (\ref{xi}) define the shadow boundary in the impact parameter space. The shadow is determined in the \textit{observer’s sky} by considering a  plane passing through the wormhole’s center and perpendicular to the line of sight, the \textit{celestial coordinates} on this plane are denoted by $\alpha$ and $\beta$ defined as \cite{Vazquez:2003zm}

\begin{equation}\label{alpha1}
\alpha = \lim_{r_o \to \infty} \left( -r_o^2 \sin \theta_o \left[ \frac{d\phi}{dr} \right]_{(r_o, \theta_o)} \right) 
\end{equation}

\begin{equation}\label{beta1}
\beta = \lim_{r_o \to \infty} \left( r_o^2 \left[ \frac{d\theta}{dr} \right]_{(r_o, \theta_o)} \right)  
\end{equation}

The observer's location is denoted by $r_o$ (in the asymptotic limit) and $\theta_o$ represents the inclination angle which is  defined by the angle between wormhole’s rotation axis and the line of sight. The derivative in the above expressions are evaluated in the asymptotic region employing the first integrals of the orbit equations. In order to describe the apparent shape of the wormhole in the presence of plasma we find the closed orbits around it considering different plasma profiles.  The apparent image of the wormhole will differ from its geometrical size, because of the refraction 
of the light rays and thus the actual cross section may differ from the geometrical one.

\subsection{Shadows for homogeneous plasma profile:} We start with the homogeneous 
plasma distribution described by the density profile in eq. (\ref{constant plasma}) where $\rho$ denotes the homogeneous plasma parameter. The magnitude of 
$\rho$ varies from 0 to 1 in order to satisfy the condition that the photons will reach the asymptotic observer without getting absorbed in their way. Employing the eqs. (\ref{constant plasma2}), (\ref{alpha1}) and (\ref{beta1}), the celestial coordinates  are found in the forms given below
\begin{eqnarray}\label{alpha hom}
\alpha (r_c)&=&-\frac{\eta \csc \theta_o
}{\sqrt{1-\rho}} \\ \nonumber
\beta(r_c)&=&\sqrt{\frac{\xi-(\eta -a)^2+a^2\cos^2\theta_o-\eta^2\cot^2\theta_o-\rho a^2 \cos^2\theta_o}{1-\rho}}
\end{eqnarray}

\begin{figure}[h]
         \includegraphics[width=0.326\textwidth]{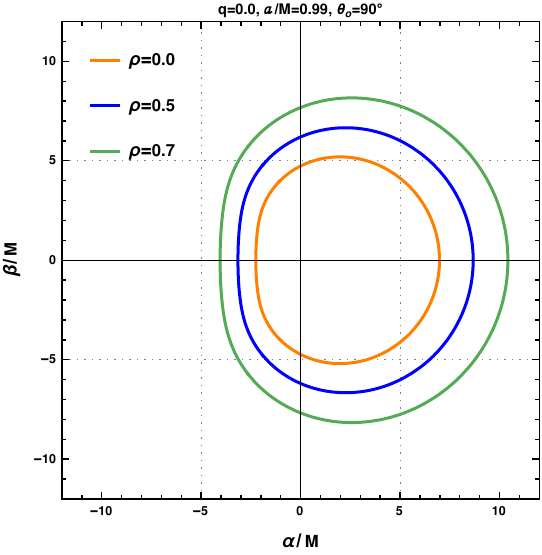}~~~
         \includegraphics[width=0.326\textwidth]{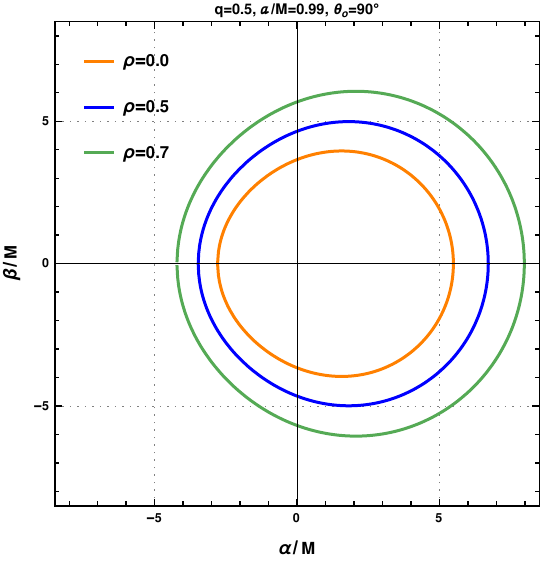}~~~
         \includegraphics[width=0.326\textwidth]{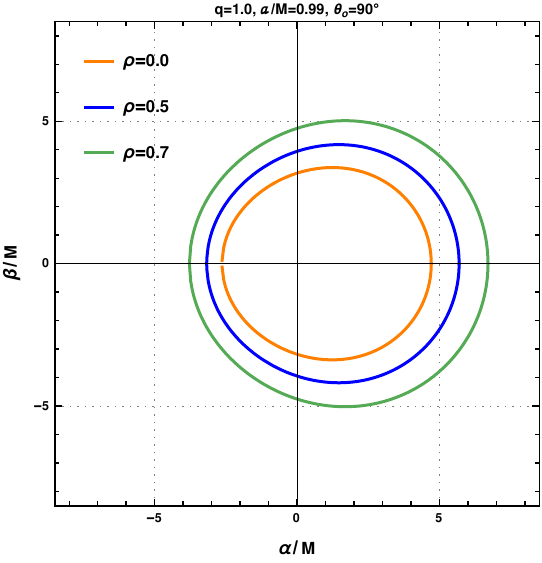}
        \caption{\textit{Shadow profile of rotating wormhole in the presence of homogeneous plasma. }}
        \label{fig:c1Whsh1}
\end{figure}

The parametric plot of $\alpha(r_c)$ and $\beta(r_c)$ is utilized to display the shadow
boundary profile in fig. (\ref{fig:c1Whsh1}) where we illustrate that the morphology of the shadow depends on the spin, angle of inclination, Kerr deviation parameter and the plasma characteristics. In fig. (\ref{fig:c1Whsh1}) we have maintained the spin parameter 
$a = 0.99$ and the angle of inclination $\theta_o = 90^\circ$. For varied plasma densities, $\rho = 0, 0.5$ and $0.7$ we have shown how the shadow profile changes with Kerr deviation parameter $q$ where the calculations assume \( \omega_0 = 1 \) and normalization with respect to \( M\).  The plots reveal that, for $q = 0$, the shadow coincides with that of the Kerr spacetime under same plasma effect, while the size of the shadow shrinks and shape alters with increasing $q$. On the other hand, for a fixed geometrical parameter $q$, an increase in the homogeneous plasma parameter enlarges the shadow appearance size. These qualitative features remain consistent for smaller spin values as well.

\begin{figure}[htb]
         \includegraphics[width=0.326\textwidth]{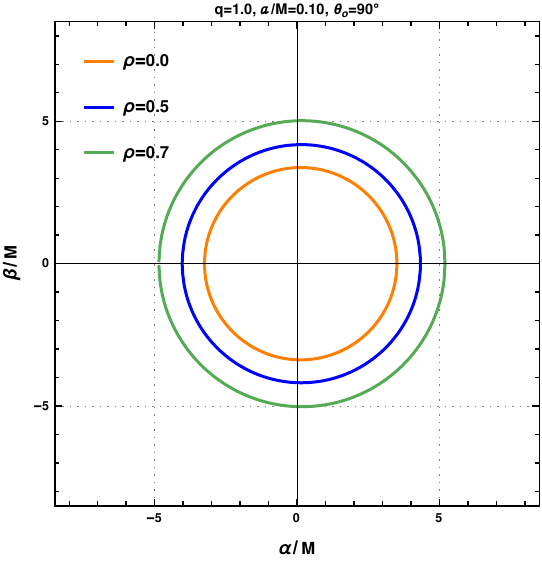}~~~
         \includegraphics[width=0.326\textwidth]{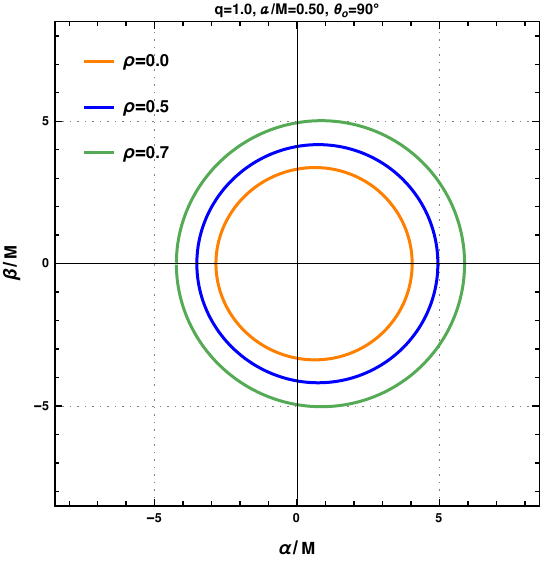}
        \caption{\textit{Wormhole shadows for \( q = 1.0 \) with different values of \( a \) and \( \rho \).}}
        \label{fig:c1Whsh2}
\end{figure}

 The influence of spin and plasma density is further illustrated in Figure \ref{fig:c1Whsh2} where the geometrical parameter $q$ is fixed. For a given density 
 parameter the shadow experiences a lateral shift towards right in the observer's sky with  increasing rotation parameter $a$.  The size change occurs for increasing density 
 parameter.

\begin{figure}[htb]       
         \includegraphics[width=0.326\textwidth]{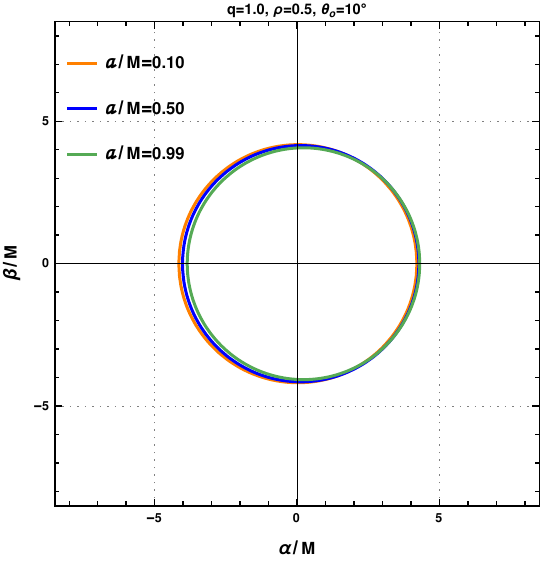}~~~
         \includegraphics[width=0.326\textwidth]{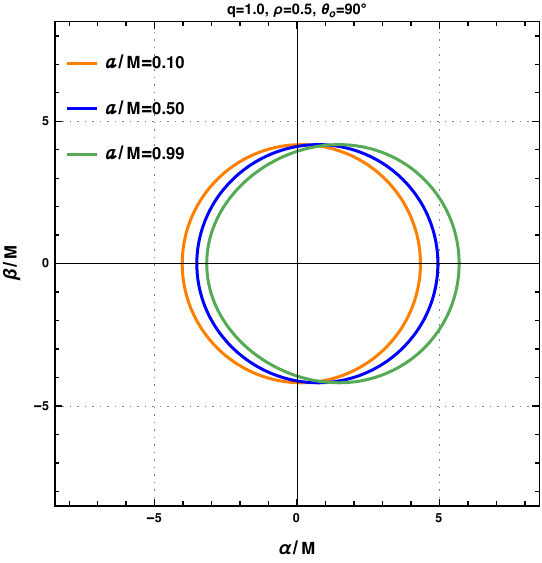}~~~
         \includegraphics[width=0.326\textwidth]{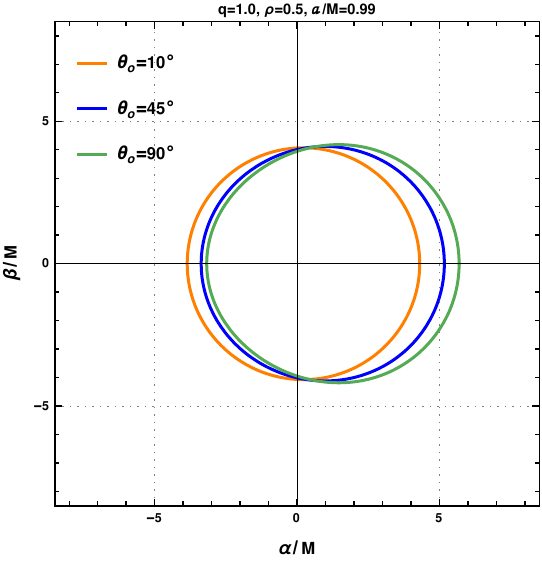}
        \caption{\textit{Shadow profiles for different \(a\) and inclination angles \(\theta_o\).}}
        \label{fig:c1Whsh3}
\end{figure}

In the first two plots in Figure \ref{fig:c1Whsh3} we show the effect of varying inclination angle $\theta_o$ and spin parameter. The plots indicate that for a given 
geometrical set up and plasma environment the shadow shifts rightward in the observer's sky and shows deviation from circularity with change in inclination angle. This effect 
gets enhanced for rapidly spinning wormhole. 

\subsection{Longitudinal plasma distribution:}

We now explore the effect of longitudinal plasma distribution where the separability 
condition leads to a polar coordinate $\theta$ dependent density profile as demonstrated 
 in eq. (\ref{plasma dist2 function}). Following the method used in the previous section 
 we find the edge of the shadow boundary by employing the eqs. (\ref{alpha1}) and (\ref{beta1}) which lead to the following expressions for the celestial coordinates 
 in this plasma distribution
\begin{eqnarray}
\alpha (r_c)&=&-{\eta \csc \theta_o} \\ \nonumber
\beta(r_c)&=&\sqrt{{\xi-(\eta -a)^2+a^2\cos^2\theta_o-\eta^2\cot^2\theta_o-\rho_\theta M^2(1+2\sin^2\theta_o)}} 
\end{eqnarray}

\begin{figure}[h]
         \includegraphics[width=0.326\textwidth]{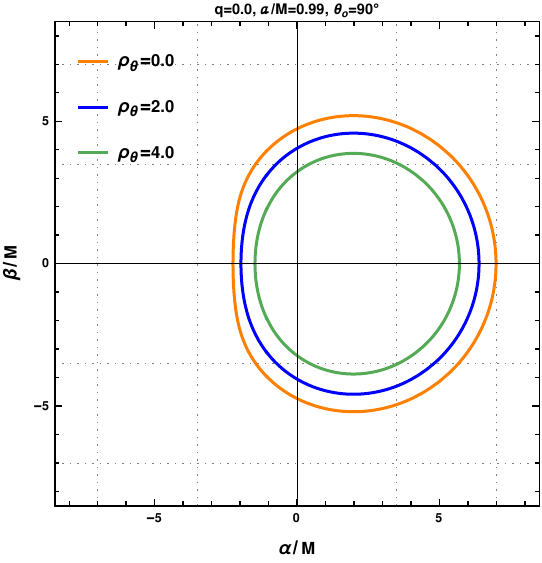}
         \includegraphics[width=0.326\textwidth]{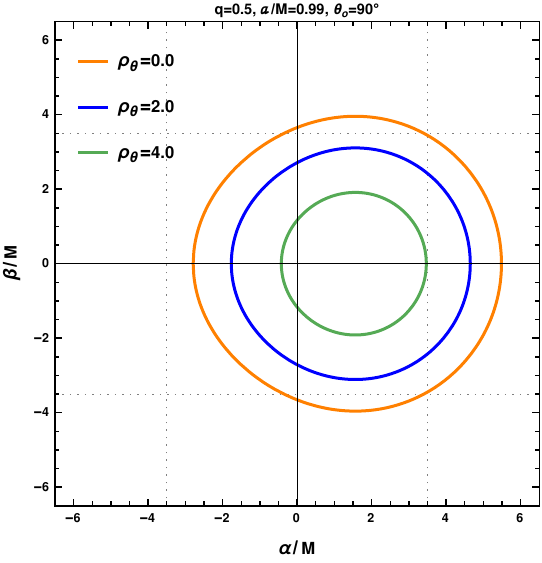}
         \includegraphics[width=0.326\textwidth]{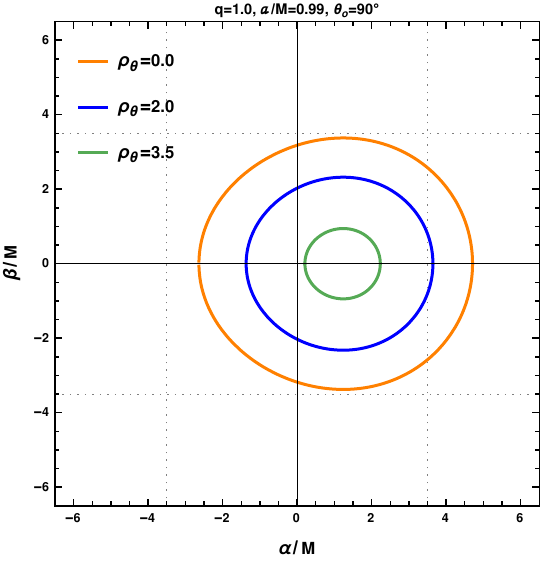}
        \caption{\textit{Shadow boundaries of the wormhole with \( a = 0.99 \) for varying \( q \) and \( \rho_\theta \).}}
        \label{fig:c2Whsh1}
\end{figure}

Comparative study of the shadow profile has been demonstrated in fig. \ref{fig:c2Whsh1} by choosing rotation parameter $a = 0.99$ and angle of inclination $\theta_o = 90^\circ$ under this particular plasma profile.  An interesting feature has been found compared 
to the previous case. The shadow size decreases monotonically with increasing $\rho_\theta$ and vanishes beyond a certain threshold, at which point the condition of 
non-absorbance of photon in the asymptotic region is no longer met. The overall dependence of the shadow on the wormhole parameters and the inclination angle for the longitudinal plasma distribution mirrors that of the homogeneous case, with the primary distinction being the monotonic shrinkage in size arising from the distinct plasma profile.  However, this finding indicates the importance of considering generalized plasma profiles over to the radial ones which are most commonly studied in the literature. 

\subsection{Radial plasma distribution:}
As our next example, we consider an inhomogeneous plasma profile where the density 
of plasma is proportional to $r^{-3/2}$ just like dust and the separability condition 
leads to the profile functions as given in eq. (\ref{plasma dist3 functions}).
The celestial coordinates corresponding to this plasma profile are given by 
\begin{eqnarray}
\alpha (r_c) &= &-{\eta \csc \theta_o} \\ \nonumber
\beta(r_c)&=&\sqrt{{\xi-(\eta -a)^2+a^2\cos^2\theta_o-\eta^2\cot^2\theta_o}}
\end{eqnarray} 

\begin{figure}[h]
         \includegraphics[width=0.326\textwidth]{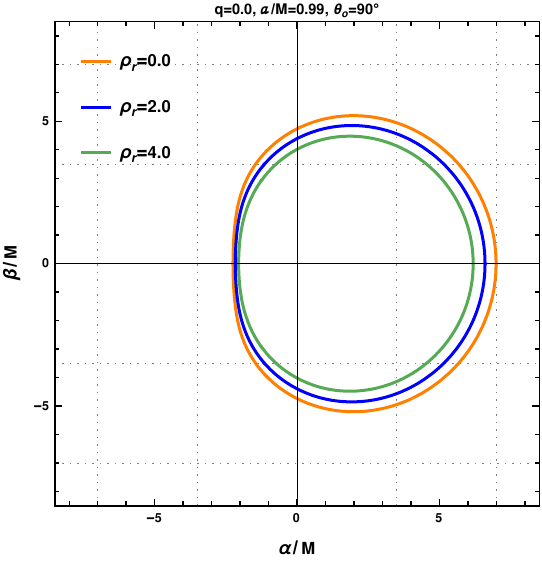}~~~
         \includegraphics[width=0.326\textwidth]{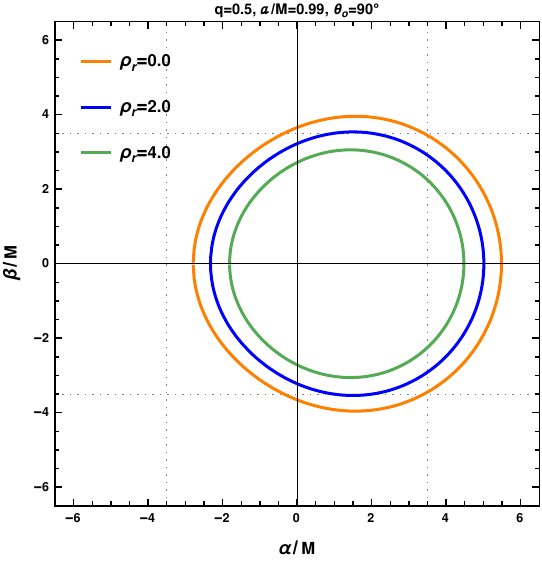}~~~
         \includegraphics[width=0.326\textwidth]{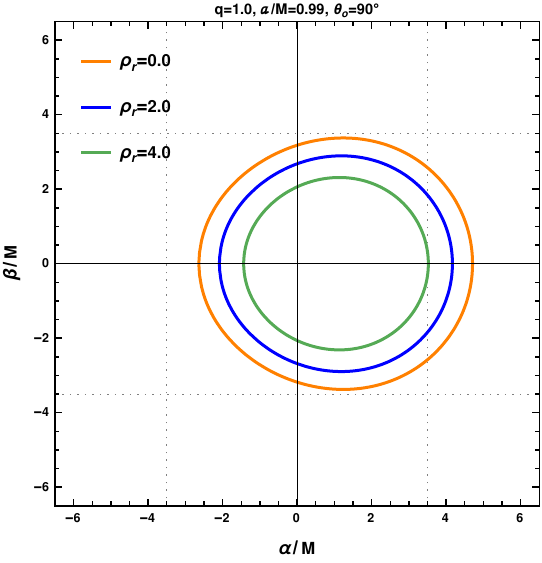}
        \caption{\textit{Shadow contours of the wormhole with \( a = 0.99 \) for different values of \( q \) and \( \rho_r \).}}
        \label{fig:c3Whsh1}
\end{figure}

Figure \ref{fig:c3Whsh1} illustrates the dependence of the shadow profile on variations in $\rho_r$ and $q$. For a fixed $q$, the shadow size decreases progressively with increasing $\rho_r$, ultimately disappearing at a particular threshold value. 
The radial plasma distribution shows a qualitatively similar effect on the wormhole shadow as observed in the longitudinal plasma case.
Since the plasma medium does not exert any direct gravitational influence on the wormhole geometry, the propagation of light in such an environment is affected both by the background spacetime and by the dispersive properties of the plasma. 
This dispersive effect leads to distinct, non-trivial variations (either enhancement or reduction) in the shadow size for different plasma profiles. 
However, variations in the wormhole parameters produce similar overall trends in the shadow behavior across different plasma environments, with the specific plasma-induced features remaining unchanged. 
These characteristic signatures in the shadow profiles are clearly seen in this work.

\section{Estimation of the Range of Geometrical and Plasma Model Parameters}{\label{Param}}
This study on the shadow of a rotating wormhole, accounting for the influence of cold and non-magnetized plasma, has been conducted primarily through analytical methods, owing to the specific form of the plasma frequency that admits Carter's constant. It is complementary to the full numerical studies which are of high relevance with observations. Although recent findings indicate the significance of considering magnetized plasma for enhanced observational relevance, in this paper we restrict the  analysis to the non-magnetized plasma model, aiming to develop a gross understanding of the parameter
space that governs both the geometrical structure and the plasma profile. 
%In principle, one can determine the parameters by fitting a calculated contour of a shadow  to an observed image.  However, practically it is impossible because there will be huge degeneracy in the parameter space which will allow shadows with a quite similar shape and size for largely different set of parameters. 
Keeping in mind that the refractive effect of plasma on photon trajectories is frequency dependent and such effects are palpable only at the radio frequency range, we employ the observational bounds obtained by the EHT for the super-massive black hole shadows of M87${}^*$ and SgrA${}^*$ \cite{EventHorizonTelescope:2019dse, EventHorizonTelescope:2022wkp} to gain a general idea about the range of parameters. Here the parameter space is spanned by the geometrical variables $M$, $a$ and $q$, along with the plasma parameters $\rho$, $\rho_\theta$ and $\rho_r$, each corresponding to one of the plasma distributions.  

For rotating objects, the effect of frame dragging on the photon geodesics breaks the circular symmetry of the shadow about the rotation axis, when compared to the static objects. The 
shape change depends on the spin and therefore, a quantity measuring the deviation from circularity, $\Delta C$, serves as a good observational tool \cite{Bambi:2019tjh} for constraining the parameter space, the spin in particular. It is apparent from the shadow profiles given above, there is a 
reflection symmetry in the  observer’s sky with respect to the $\alpha$-axis (\textit{i.e.}, $\beta=0$).
The geometric center, ($\alpha_{gc}$, $\beta_{gc}$), of the shadow is determined by calculating the average of $\alpha$ across the full shadow profile: $\alpha_{gc} ={\int \alpha \, dA}/{\int dA}, ~ \beta_{gc} = 0$ where $dA = 2\beta d\alpha$ denotes the area element of the shadow profile. To quantify the shift of the shadow let us consider an arbitrary 
point, $Q (\alpha(\psi), \beta(\psi))$, on the boundary of the shadow, where $\psi$ is the angle between 
the $\alpha$-axis and the vector joining the point $Q$ and the geometric center of the shadow. The corresponding radial distance $\ell(\psi)$ is given by $ \sqrt{\left(\alpha (\psi)-\alpha_{gc}\right)^2-\beta (\psi)^2}
$ where $\psi= \tan ^{-1}\left\{\beta(\psi)/(\alpha (\psi)-\alpha_{gc})\right\}$. Employing this expression we define the average shadow radius as 
\begin{equation}
    \bar{R}^2 = \frac{1}{2\pi} \int_0^{2\pi}   \ell^2(\psi)\, d\psi
\end{equation}
The fractional deviation parameter quantifying the shift in shadow due to strong frame dragging effect thus can be obtained following the prescription given in \cite{EventHorizonTelescope:2019dse} where 
it is defined as the fractional root mean square distance from the average shadow radius ($\bar{R}$) 

\begin{equation} \label{eq:circular dev}
    \Delta C = \frac{1}{\bar{R}} \sqrt{\frac{1}{2\pi} \int_0^{2\pi} (\ell(\psi) - \bar{R})^2 \, d\psi}
\end{equation}

In the spacetime under consideration there are three geometrical parameters, $a/M$ and $q$, corresponding to each plasma configuration. The deviation of the shadow from circularity strongly depends on spin parameter. As EHT observation could not give any bound on shadow circularity or the spin for SgrA${}^*$, the allowed range of the parameters, $a/M$ and $q$ for a given plasma configuration, may be obtained using the  M87${}^*$ observational bounds 
%where the distance of the compact object from earth, $D = 16.8~\mbox{Mpc}$, estimated mass range, 
%$M = (6.5 \pm 0.7) \times 10^{7} M_{\odot}$, 
corresponding angle of inclination,  $\theta_o = 17^{\circ}$ which gives flux depression in the central emission region of 
angular diameter $42 \pm 3 ~\mbox{$\mu$as}$ by a factor of 10 and the deviation from circularity $\Delta C \leq 0.10$ \cite{EventHorizonTelescope:2019dse, EventHorizonTelescope:2019pgp, EventHorizonTelescope:2019ggy}. Further we constrain the plasma parameter by using the bound on emission ring size within the restricted range of the geometrical parameters thus achieved.\\

\begin{figure}[h]
         \includegraphics[width=0.326\textwidth]{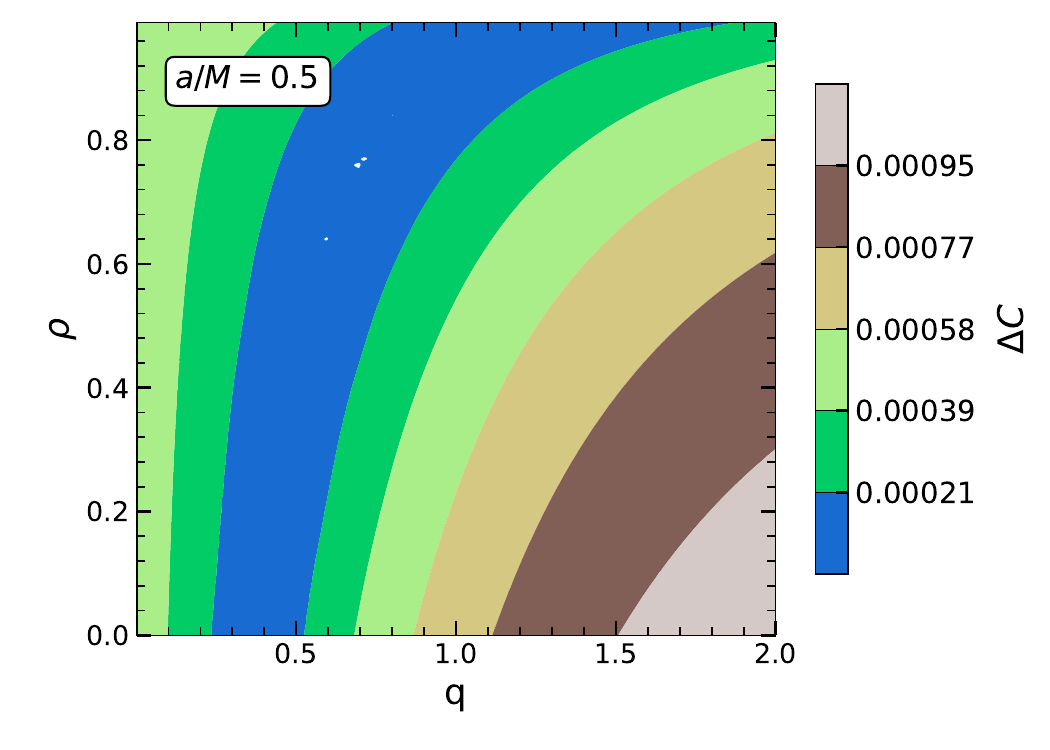}
         \includegraphics[width=0.326\textwidth]{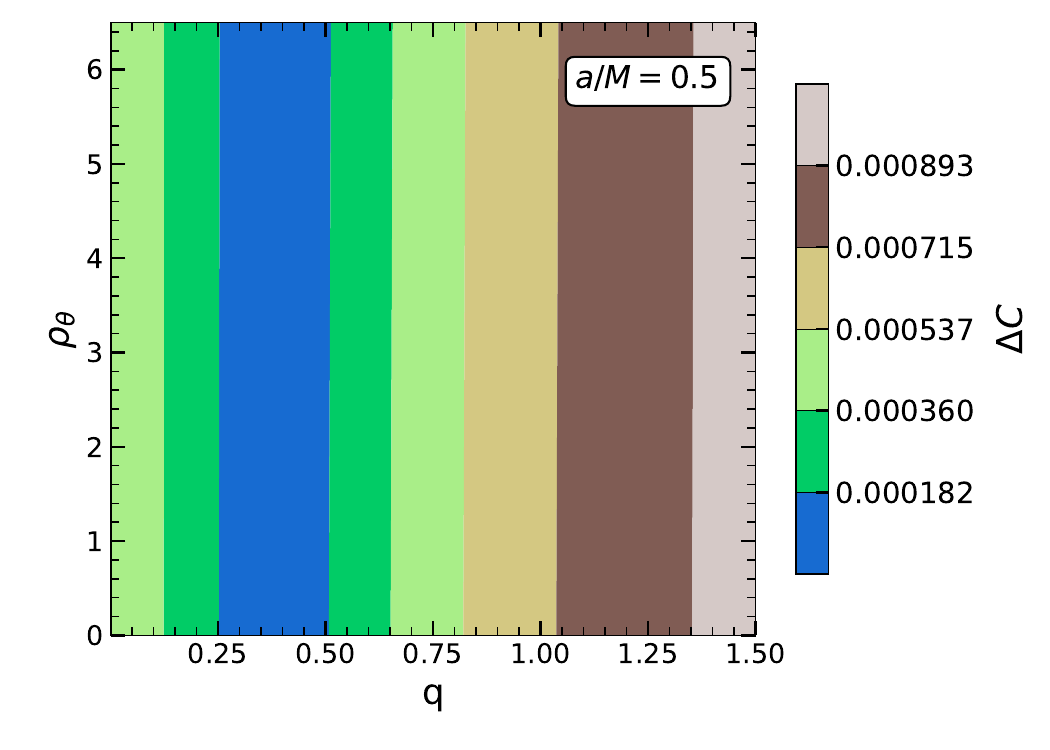}
         \includegraphics[width=0.326\textwidth]{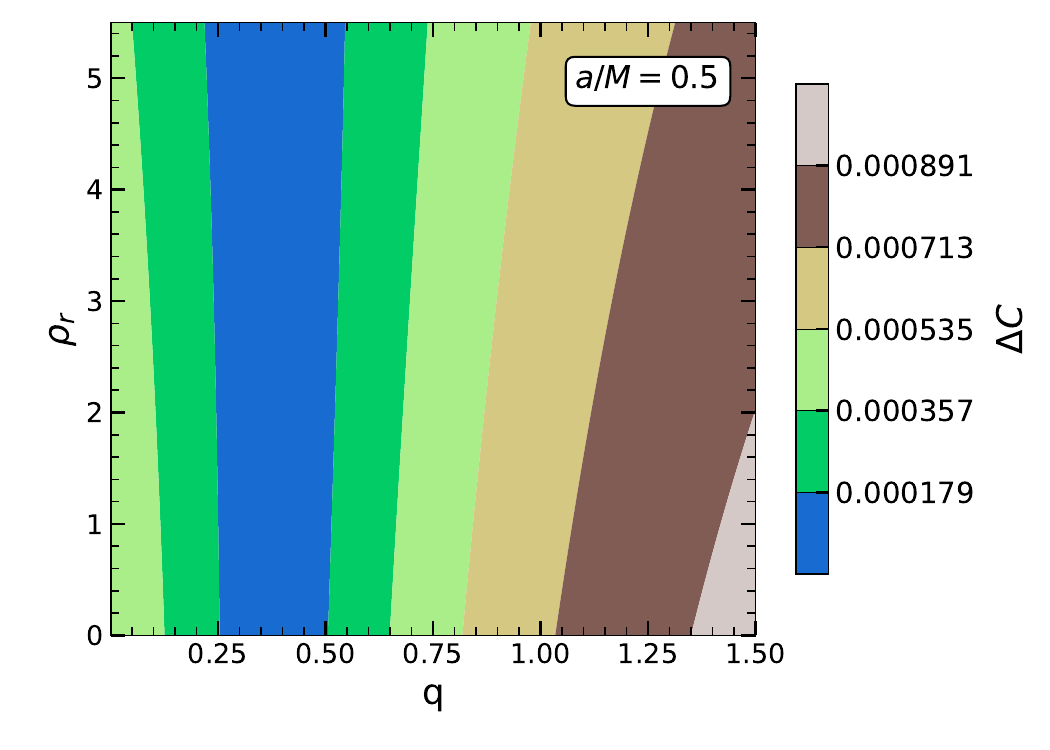}
        \caption{\textit{ Deviation from circularity for homogeneous(left), longitudinal(middle) and radial(right) plasma distribution}}
        \label{dev_cir_hom}
\end{figure}

\begin{figure}[htb]
         \includegraphics[width=0.326\textwidth]{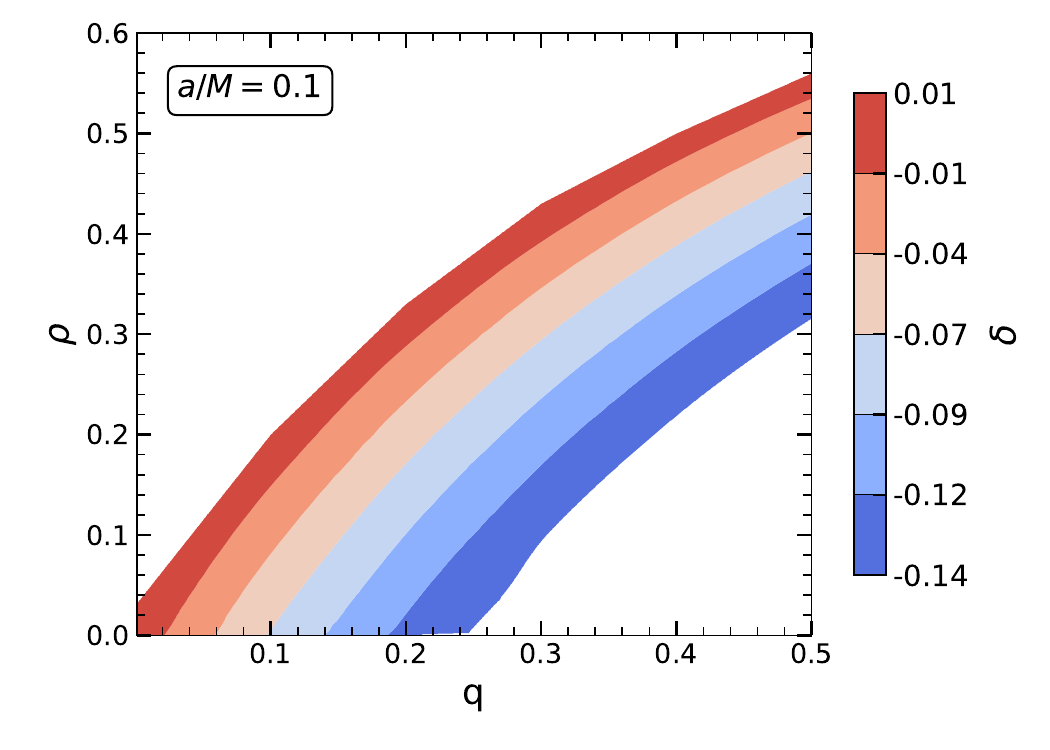}~~~    
         \includegraphics[width=0.326\textwidth]{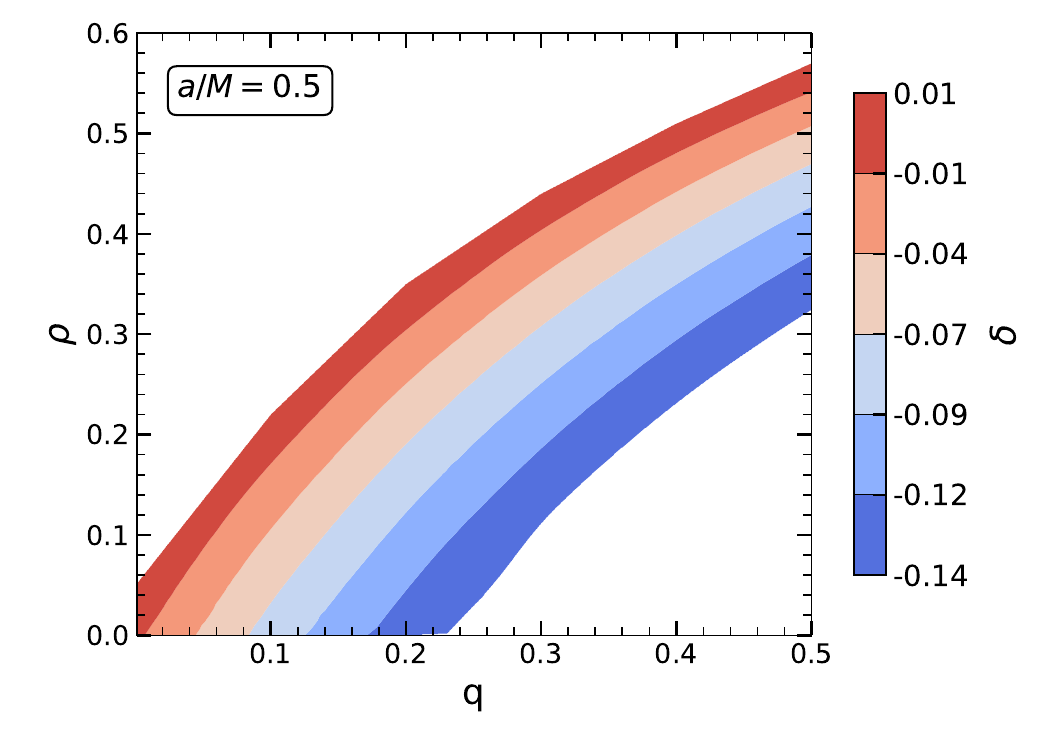}~~~
         \includegraphics[width=0.326\textwidth]{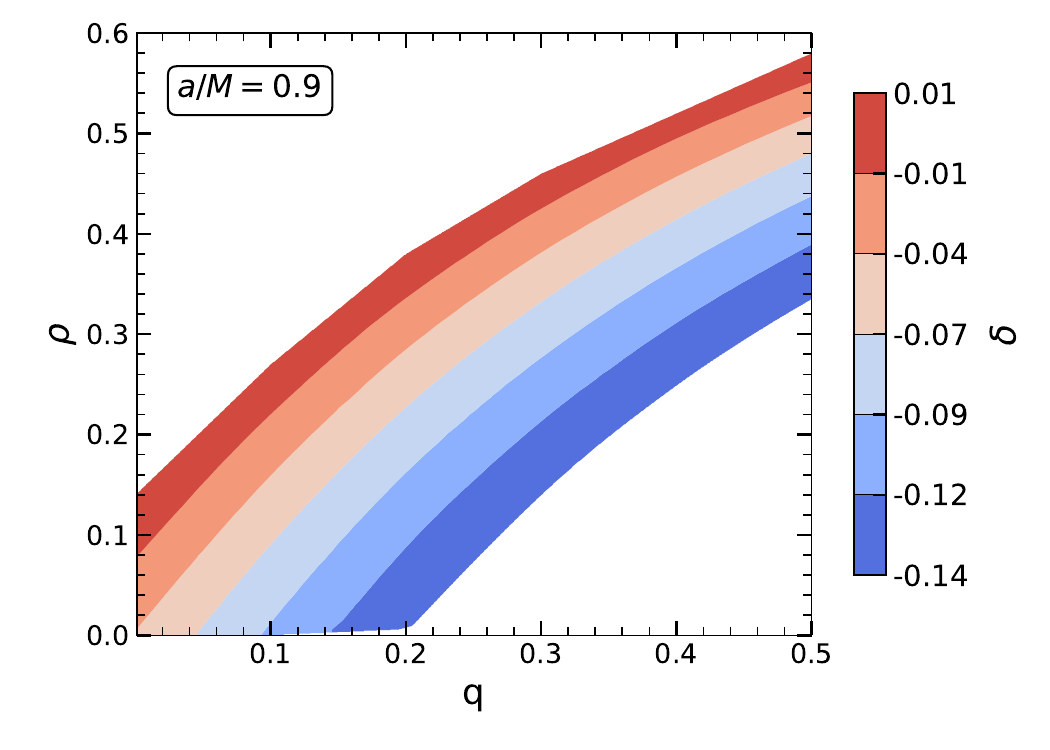} 
        \caption{\textit{Fractional deviation parameter and the allowed span of $q$ and $\rho$ for various $a/M$.}}
        \label{frac_dev_hom_af}
\end{figure}

In fig. \ref{dev_cir_hom}, we have shown the parameter spaces spanning over $q$ and $\rho$ and the associated range of $\Delta C$ for the homogeneous, longitudinal and radial  plasma environments. For homogeneous plasma environment, we have found for the theoretically bounded 
values of $\rho$ the observational bound on circular deviation parameter is satisfied for 
entire allowed value of $q$ and thus no definitive additional bound on $q$ or plasma profile can be established by using this parameter. The same applies to longitudinal and radial plasma environment.
All of these findings are also true for the entire range of spin parameter. Thus we move 
on to exploring the bounds on fractional deviation parameter ($\delta$) characterizing the deviation of the average shadow diameter inferred by the geometry from that of the Schwarzschild BH \cite{EventHorizonTelescope:2022xqj} which is defined by

\begin{equation} \label{eq:frac dev para}
    \delta=\left(\frac{\theta_{sh}}{\theta_{sh, Sch}}-1\right)=\left(\frac{\bar{R}}{3\sqrt{3}M}-1 \right) 
\end{equation}

where average diameter of the shadow is  $2\bar{R}$. The 
parameters will be constrained by using the EHT measured shadow diameter of  SgrA${}^*$ ($\theta_{sh} = (48.7 \pm 7) \mu as$) and the range of fractional deviation parameter $-0.14 < \delta < 0.01$, which lies within the observational limits reported by both VLTI and Keck \cite{EventHorizonTelescope:2022xqj}. To define this range 
EHT used two separate sets of mass and distance estimates obtained from VLTI and Keck and also employed three independent imaging algorithms  \cite{EventHorizonTelescope:2022wkp, EventHorizonTelescope:2022xqj}. In the Figure \ref{frac_dev_hom_af}, we have shown the span of the parameters $q$ and $\rho$ satisfying the allowed 
range of $\delta$ for homogeneous plasma profile. This figure demonstrates how one should pick the values of $q$ for a given plasma density profile. 

\begin{figure}[h]      
         \includegraphics[width=0.326\textwidth]{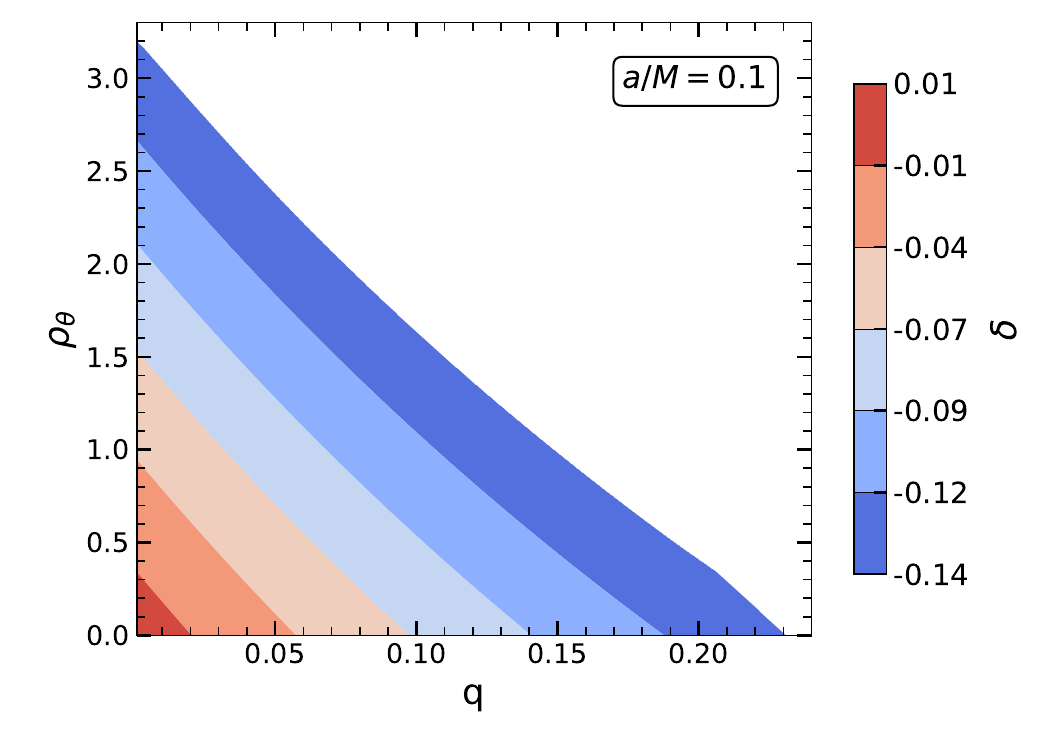}~~~
         \includegraphics[width=0.326\textwidth]{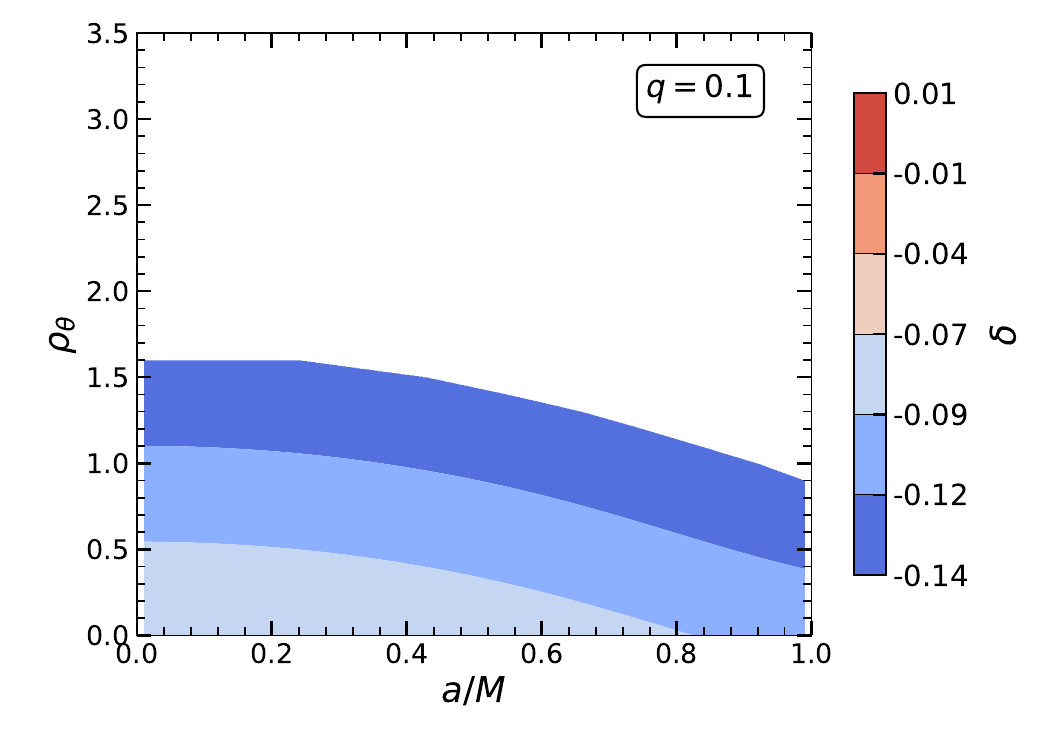}~~~
         \includegraphics[width=0.326\textwidth]{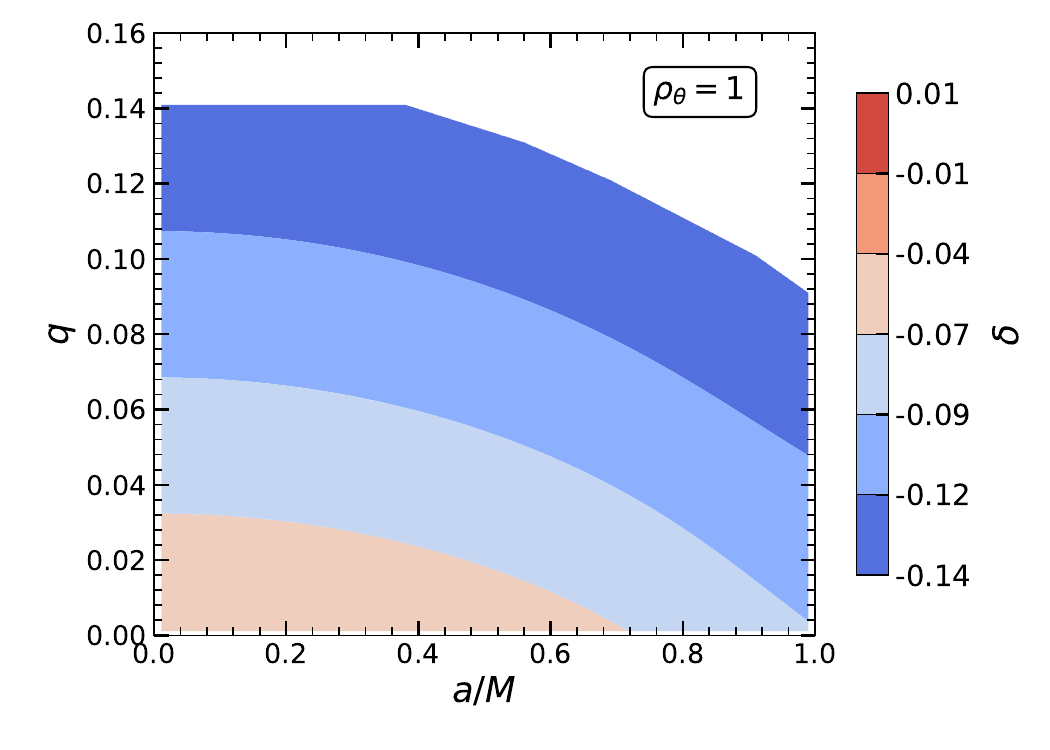}
        \caption{\textit{Parameter ranges corresponding to longitudinal plasma profile.}}
        \label{frac_dev_long_af}
\end{figure}

The first panel in figure (\ref{frac_dev_long_af}) depicts the range of coupled parameter space spanned over 
$q$ and $\rho_\theta$ for longitudinal plasma profile for given spin. Note that, an upper bound on $q \sim 0.23$ 
is obtained for $\rho_{\theta} \sim 3.0$ corresponding to $a/M = 0.1$. The parameter space of $q$ and $\rho_\theta$ shrinks with increasing of spin parameter $a/M$ in order to maintain the consistency with observational bound by  
EHT result for SgrA${}^*$. The middle panel illustrates the parameter space in the $\rho_\theta$ - $a/M$ plane and the $q$ - $a/M$ parameter space is depicted in the right panel. By considering the entire theoretically allowed range of parameters compatible with the observational constraint, we obtain the following upper limits on the parameters 
in a longitudinal plasma environment: \( q \approx 0.24 \), \( a/M \approx 0.99 \), and \( \rho_{\theta} \approx 3.2 \). Similar analysis in the case of radial plasma has been performed. In this case the parameter space of $q$ - $\rho_r$ gets squeezed with the increasing spin.  Our study on shadow in radial plasma environment suggests the following bounds on the parameters: \( q \approx 0.24 \), \( a/M \approx 0.99 \), and \( \rho_r \approx 4 \). 

\section{Summary and Future Directions}

In the current work we explored interesting features in the shadow properties of a horizon less object by studying the null trajectories in the presence of plasma. We have revisited the features of  a newly invented rotating wormhole geometry (Eq. \ref{rotating wh}) which has been originated from a static Ricci flat wormhole  spacetime and has the possibility of being sourced by energy condition non-violating matter in the higher dimensional scenario \cite{Kar:2015lma, Kar:2024ctd}. Our investigation was focused on exploring the light trajectories in the presence of a more astrophysics wise relevant environment of non-magnetized and cold plasma models with  three distinct (homogeneous, longitudinal and radial) plasma density profiles. The direct influence of plasma energy-momentum on the geometry has been disregarded whereas the emphasis has been given on the refracting properties of plasma, in particular to the cases where  turning points are  there in the photon trajectories. We have employed the modified Hamiltonian in order to incorporate the plasma energy density profile obtained from the Drude-Lorentz model. Finally to find the photon trajectories we apply the Hamilton-Jacobi 
formalism and a key feature of this rotating wormhole spacetime is that it allows separation of variables. The existence of Carter's constant has been achieved by judiciously choosing the 
forms of the plasma density profile. This property allows the shadow boundary to be computed  
analytically by interpreting it as a gravitationally lensed projection of unstable spherical orbits. \\

% Shadow morphology

Shadow morphology encodes influences of both the spacetime parameters and the properties of plasma medium. In section \ref{shadow study} we have shown for homogeneous plasma the shadow coincides with that of the Kerr for $q = 0$ and remarkably 
the size of the shadow shrinks and shape alters with increasing values of the deviation parameter $q$. On the other hand an increase in the homogeneous plasma parameter enlarges the shadow appearance size. These two parameters play a contrasting role to the shadow morphology. On the other hand, for a given geometrical set up and plasma environment the shadow shifts rightward in the observer’s sky and deviation from circularity increases with inclination angle which gets further enhanced for rapidly spinning wormhole. We have found a \textit{remarkable feature} for the longitudinal plasma profile -- the monotonic shrinkage in size of the shadow with increasing $\rho_{\theta}$ and vanishing beyond a certain threshold. This finding indicates the importance of considering generalized plasma profiles.    
% Parameter Estimation
Subsequently in section \ref{Param}, we attempt to acquire a general understanding about the space spanned over the 
geometrical parameters and the plasma density parameters by utilizing the observational bounds from EHT observations of M87$^{*}$ and SgrA$^{*}$ on  deviation from circularity and the fractional deviation parameter. We have done 
the estimation by keeping one of the three parameters (out of $a/M$, $q$, $\rho$) fixed and varied the other two 
to find the allowed range within the limit of observed $\Delta C$ and $\delta$.
We derive the following bounds on the parameter space through our analysis:  \( q \approx 0.24 \), \( a/M \approx 0.99 \), and \( \rho_{\theta} \approx 3.2 \) for the longitudinal and for radial plasma environment \( q \approx 0.24 \), \( a/M \approx 0.99 \), and \( \rho_r \approx 4 \).  For homogeneous plasma profile an upper bound on the plasma density parameter $\rho \approx 1$ can be found, however we could not propose any limit on the parameter $q$.

% Future Directions
The intensity contrast in the inside and the outside of the photon ring will serve as an important 
distinguishing feature, which we plan to investigate in our future project. Incorporating magnetized plasma and a full MHD accretion flow will underscore even more realistic astrophysical environments. In computing the shadow, we assume that the wormhole is observed against a background of light sources distributed in its vicinity, with no emitters located along the line of sight between the observer and the wormhole. However, it will be another interesting project to explore the scenario when sources are there in the other side of the throat. The weak and strong lensing of light in the rotating wormhole spacetime under consideration may provide a potential tool to test  the nature of the compact object, by comparing the resulting lensing signatures with those associated with the black holes.

\section*{Acknowledgements}
P. Gayen acknowledges the financial support received from the University Grants Commission (UGC), India, through the fellowship with ID: 191620137660.

\medskip
\nocite{*}
\bibliography{Referencesv1}

\end{document}